\documentclass[prl,twocolumn,twoside,preprintnumbers,superscriptaddress,nofootinbib]{revtex4}

\usepackage{amsmath,slashed}
\usepackage{graphicx,graphics,color}
\usepackage{dcolumn}
\usepackage[hyperfootnotes=false]{hyperref}
\usepackage{xspace}
\usepackage{enumerate}
\usepackage{varwidth}
\usepackage{comment}
\usepackage{physics}

\usepackage{amssymb}
\usepackage{mathtools}
\usepackage{dsfont}
\usepackage{multirow}
\usepackage{bbold}
\usepackage{xfrac}

\usepackage{cancel}
\usepackage{orcidlink}

\newcommand{\beq}{\begin{equation}}
\newcommand{\eeq}{\end{equation}}

\newcommand{\bea}{\begin{eqnarray}}
\newcommand{\eea}{\end{eqnarray}}
\newcommand{\be}{\begin{equation}}
\newcommand{\ee}{\end{equation}}
\newcommand{\bma}{\begin{pmatrix}}
\newcommand{\ema}{\end{pmatrix}}

\newcommand{\mupi}{\mu_\pi}

\newcommand{\intem}{1.8/2.0~(EM)}
\newcommand{\intsat}{$\rm NNLO_{sat}$}

\allowdisplaybreaks[1]

\begin{document}

\preprint{INT-PUB-26-044
}

\title{Determination of \texorpdfstring{$\boldsymbol{V_{ud}}$}{} from superallowed \texorpdfstring{$\boldsymbol{\beta}$}{} decays using
effective field theory}

\author{Samuel Novario\,\orcidlink{0000-0001-7836-1269}}
\affiliation{Department of Physics, Washington University in Saint Louis, Saint Louis, MO 63130, USA}
\affiliation{McDonnell Center for the Space Sciences at Washington University in St. Louis, MO 63130, USA}

\author{Graham Chambers-Wall\,\orcidlink{0000-0002-9403-8450}}
\affiliation{Department of Physics, Washington University in Saint Louis, Saint Louis, MO 63130, USA}

\author{Vincenzo Cirigliano\,\orcidlink{0000-0002-9056-754X}}
\affiliation{Institute for Nuclear Theory, University of Washington, Seattle WA 91195-1550, USA}

\author{Wouter Dekens\,\orcidlink{0000-0002-7850-5901}}
\affiliation{Institute for Nuclear Theory, University of Washington, Seattle WA 91195-1550, USA}

\author{Jordy~de~Vries\,\orcidlink{0000-0001-5037-5108}}
\affiliation{Institute for Theoretical Physics Amsterdam and Delta Institute for Theoretical
Physics, University of Amsterdam, Science Park 904, 1098 XH Amsterdam, The Netherlands}
\affiliation{Nikhef, Theory Group, Science Park 105, 1098 XG, Amsterdam, The Netherlands}

\author{Stefano Gandolfi\,\orcidlink{0000-0002-0430-9035}}
\affiliation{Theoretical Division, Los Alamos National Laboratory, Los Alamos, NM 87545, USA}

\author{Martin Hoferichter\,\orcidlink{0000-0003-1113-9377}}
\affiliation{Albert Einstein Center for Fundamental Physics, Institute for Theoretical Physics, University of Bern, Sidlerstrasse 5, 3012 Bern, Switzerland}

\author{Emanuele Mereghetti\,\orcidlink{0000-0002-8623-5796}}
\affiliation{Theoretical Division, Los Alamos National Laboratory, Los Alamos, NM 87545, USA}

\begin{abstract}
A precision determination of $V_{ud}$ from superallowed $\beta$ decays requires exquisite control over radiative corrections, including nuclear-structure-dependent effects whose evaluation has traditionally relied on nuclear models. To assess the uncertainties inherent in the nuclear matrix elements,
an alternative strategy based on effective field theory (EFT) has been established in recent years.
Here, we present material progress in this program, in terms of a first EFT calculation of the crucial correction $\bar\delta_\text{NS}$ for 12 superallowed transitions ranging from  $^{10}$C $\rightarrow$ $^{10}$B to $^{54}$Co $\rightarrow$ $^{54}$Fe. These main results rely on
nuclear ab initio calculations using the coupled cluster method, benchmarked against Quantum Monte Carlo evaluations for light nuclei. The global fit to superallowed decay rates reveals that the precision in $V_{ud}$ is currently limited by short-distance contributions that in the EFT are parameterized as two-nucleon contact interactions, leading to an uncertainty $\delta V_{ud}\simeq 6\times 10^{-4}$. We discuss the consequences for the status of the first-row unitarity test of the CKM matrix as well as a roadmap how to reduce the uncertainty in $V_{ud}$ in the future.
\end{abstract}

\maketitle

\emph{Introduction}---For several years there have been indications for a tension in the first-row unitarity test of the Cabibbo--Kobayashi--Maskawa (CKM) matrix~\cite{Cabibbo:1963yz,Kobayashi:1973fv}
\begin{align}
\label{unitarity}
 |V_{ud}|^2+|V_{us}|^2+|V_{ub}|^2=1\,,
\end{align}
e.g., in the evaluation of Ref.~\cite{Cirigliano:2022yyo} the global fit suggests a deficit of $2.8\sigma$. This tension is partly driven by the determination of $V_{ud}$ from superallowed $\beta$ decays~\cite{Hardy:2020qwl,Gorchtein:2023naa,Hayen:2017pwg}---nuclear $0^+\to 0^+$ transitions that are exclusively sensitive to the vector component of the electroweak (EW) current---especially, its small uncertainty at the level of $\delta V_{ud}\simeq 3\times 10^{-4}$, e.g., Ref.~\cite{Hardy:2020qwl} quotes
\begin{equation}
\label{Vud_HT}
V_{ud}^{0+}=0.97373(31)\,.
\end{equation}
Arriving at such a precision requires exquisite control over radiative corrections, most critically corrections that depend on the nuclear structure of a given transition such as $\bar\delta_\text{NS}$~\cite{Seng:2018qru,Gorchtein:2018fxl,Seng:2022cnq}. Traditionally, the required nuclear matrix elements (NMEs) have been evaluated using nuclear models~\cite{Kinoshita:1958ru,Sirlin:1967zz,Sirlin:1967zza,Abers:1968zz,Jaus:1970tah,Sirlin:1977sv,Sirlin:1981ie,Wilkinson:1982hu,Sirlin:1986cc,Towner:1992xm,
Wilkinson:1993hxz,Wilkinson:1993fva,Czarnecki:2004cw,Marciano:2005ec,Towner:2010zz,Hardy:2014qxa}, posing a challenge for a robust uncertainty quantification. In fact, Eq.~\eqref{Vud_HT} already involves an increased uncertainty compared to earlier evaluations $V_{ud}= 0.97417(21)$~\cite{Hardy:2014qxa} due to nuclear corrections as estimated in Refs.~\cite{Seng:2018qru,Gorchtein:2018fxl}, while the shift in central value arises primarily from the universal ``$\gamma W$ box'' correction~\cite{Marciano:2005ec,Seng:2018yzq,Seng:2018qru,Czarnecki:2019mwq,Seng:2020wjq,Hayen:2020cxh,Shiells:2020fqp,Cirigliano:2022yyo,Ma:2023kfr}. Given the implications of a unitarity deficit for physics beyond the Standard Model~\cite{Belfatto:2019swo,Coutinho:2019aiy,Cheung:2020vqm,Belfatto:2021jhf,Branco:2021vhs,Crivellin:2021bkd,Crivellin:2020ebi,Kirk:2020wdk,Crivellin:2021njn,Crivellin:2020lzu,Crivellin:2020klg,Capdevila:2020rrl,Crivellin:2021sff,Crivellin:2020oup,Marzocca:2021azj,Alok:2021ydy, Cirigliano:2022qdm,Cirigliano:2023nol,Dawid:2024wmp}, it is of utmost importance to obtain a reliable error in $V_{ud}$ that adequately accounts for nuclear uncertainties. To address this objective systematically, we proposed a formalism rooted in effective field theory (EFT) in Refs.~\cite{Cirigliano:2024rfk,Cirigliano:2024msg}, allowing one, at each step of the calculation, to exploit the respective hierarchy of scales and assign corresponding uncertainty estimates. In this Letter, we report material progress in this program, with a detailed account provided in the companion paper~\cite{companion}.

The first step in the EFT concerns the renormalization group (RG) evolution of the semileptonic operator mediating $\beta$ decay from the EW scale down to the nuclear scale. With the perturbative evolution known at next-to-leading-logarithmic order~\cite{Cirigliano:2023fnz} including $\alpha_s$ corrections~\cite{Moretti:2025ghw}, the overall precision is determined by that of the NMEs, allowing, in principle, for $V_{ud}$ determinations below $1\times 10^{-4}$, as demonstrated by the extraction from pion $\beta$ decay~\cite{Cirigliano:2026ios}
\beq
\label{Vud_pion}
V_{ud}^\pi=0.97346(281)_\text{Br}(9)_{\tau_\pi}(27)_{\Delta_\pi}(5)_\text{th}[283]_\text{tot}\,.
\eeq
In this case, thanks to nonperturbative input from lattice QCD~\cite{Feng:2020zdc,Yoo:2023gln}, the remaining theory uncertainty (th) is dwarfed by the experimental uncertainties due to pion lifetime ($\tau_\pi$), mass difference ($\Delta_\pi$), and branching fraction (Br), bringing a precision as quoted in Eq.~\eqref{Vud_HT} within reach once the latter improves accordingly~\cite{PIONEER:2022yag,PIONEER:2025idw}.

Matrix elements at the single-nucleon level are already less well determined, and the nonperturbative uncertainty related to the $\gamma W$ box contribution~\cite{Marciano:2005ec,Seng:2018yzq,Seng:2018qru,Czarnecki:2019mwq,Seng:2020wjq,Hayen:2020cxh,Shiells:2020fqp,Cirigliano:2022yyo,Ma:2023kfr} propagates to the $V_{ud}$ determination from superallowed $\beta$ decays in terms of the effective Wilson coefficient $g_V$ at the low scale.
Combined with improved perturbative corrections~\cite{Hill:2023acw,Hill:2023bfh,Borah:2024ghn,VanderGriend:2025mdc,Cao:2025zxs,Crosas:2025xyv} and using only the best individual measurements for neutron lifetime $\tau_n$~\cite{UCNt:2021pcg} and asymmetry $\lambda$~\cite{Markisch:2018ndu}, one obtains
\beq
\label{Vud_neutron}
V_{ud}^{n,\,\text{best}}=0.97403(20)_{\tau_n}(35)_{\lambda}(2)_{\Delta_N}(13)_\text{th}[42]_\text{tot}\,,
\eeq
where we used the $\gamma W$ box evaluation from Refs.~\cite{Cirigliano:2022yyo,Cirigliano:2023fnz}. This uncertainty already comes tantalizingly close to Eq.~\eqref{Vud_HT}, but increases by more than a factor of two when the scale factors in $\tau_n$ and $\lambda$ are taken into account~\cite{ParticleDataGroup:2026mpi}.
For superallowed $\beta$ decays the experimental inputs would allow for a determination at a level of precision similar to the uncertainty in $g_V$, which almost coincides with the theory error in Eq.~\eqref{Vud_neutron}, but the crucial question becomes how much radiative corrections sensitive to NMEs increase this best-case uncertainty $\delta V_{ud}\lesssim 2\times 10^{-4}$. In this Letter we report which uncertainty in $V_{ud}$ we believe can currently be justified from the EFT perspective.

\emph{Half-life and nuclear matrix elements}---In the EFT formalism~\cite{Cirigliano:2024rfk,Cirigliano:2024msg} the superallowed half-life is expressed as
\begin{align}
    \frac{1}{t}  &=    \frac{G_F^2 |V_{ud}|^2  m_e^5}{\pi^3 \log 2}\,
    \Big[C_\text{eff}^{(g_V)}(\mu)\Big]^2  \notag\\
    &\quad\times[1 + \bar \delta_R^\prime(\mu)]\,  (1 +  \bar \delta_\text{NS}) \,  ( 1 - \bar  \delta_C)  \,  \bar f(\mu)\,,
    \label{eq:master}
\end{align}
where $\mu$ denotes the renormalization scale, and the product of the Wilson coefficient $C_\text{eff}^{(g_V)}(\mu)$ squared and the factors $[1 + \bar \delta_R^\prime(\mu)]\bar f(\mu)$ is scale independent up to higher perturbative orders. [The bar indicates definition in the EFT, to distinguish the notation from the traditional approach~\cite{Hardy:2020qwl}.] $C_\text{eff}^{(g_V)}(\mu)$ encodes the effects of hard photons, including the RG from the EW to the nucleon scale, the nonperturbative matching involving the $\gamma W$ box, and the perturbative QED evolution to the nuclear physics scale, while $[1 + \bar \delta_R^\prime(\mu)]\bar f(\mu)$ receives contributions from ultrasoft photons, including effects from low-energy scales such as the ${\mathcal Q}_\text{EC}$ value of the reaction. Finally, potential and soft photons contribute to $\bar\delta_C$ and $\bar\delta_\text{NS}$, which, to leading order in the fine-structure constant $\alpha=e^2/(4\pi)$, correspond to isospin-breaking corrections in the nuclear wave function and two-body corrections to the EW current, respectively. While ab initio calculations of $\bar\delta_C$ are still being developed~\cite{Miller:2008my,Miller:2009cg,Seng:2023cvt,Farren:2024spl,Piarulli:2026abq,Seng:2026hwk}, in this Letter we present a first EFT evaluation of $\bar\delta_\text{NS}$.

Decomposing this correction into energy-independent and -dependent components
\begin{equation}\label{eq:deltaNS}
    \bar\delta_\text{NS} = \delta_\text{NS}^{(0)} + \overline{\delta_\text{NS}^{E}}\,,
\end{equation}
and including terms up to 
$\mathcal O(\alpha \epsilon_\chi)$ 
and $\mathcal O(\alpha^2)$ in the expansion in chiral EFT ($\epsilon_\chi$ is the chiral expansion parameter~\cite{Cirigliano:2024msg}),
\begin{align}\label{deltaNS0}
   \delta^{(0)}_\text{NS}  &= \frac{2}{g_V(\mupi)M_\text{F}^{(0)}}   \sum_{N = n,p} \bigg[ \alpha  \Big( M^\text{mag}_{\text{GT}, N} + M^\text{mag}_{\text{T}, N} +  M^\text{CT}_{\text{GT}, N}   \nonumber \\ &  + M_{\text{LS}, N}  \Big)
    + g_V(\mu_\pi)\alpha^2 \Big( M^+_{\text{F}, N} +
    M^{+\, \text{3b}}_{\text{F}, N} \Big)  \bigg]\,,\notag\\
 \overline{\delta^E_\text{NS}} &= - \alpha \frac{2}{M_\text{F}^{(0)}} R_A E_0
\tilde{f}_E  M^E_{\text{F}, p}\,,
\end{align}
we require the Gamow--Teller (GT) NMEs, corresponding contact-term NMEs needed for their renormalization, tensor (T) and spin-orbit (LS) NMEs, and $\alpha^2$ corrections to Fermi (F) NMEs, normalized to $M_\text{F}^{(0)}=\sqrt{2}$ and $g_V$ at the scale $\mu_\pi=\sqrt{\frac{3}{5}} R^{-1}_\text{ch}e^{1/2-\gamma_E}$, with nuclear charge radius $R_\text{ch}$. $E_0$ denotes the positron endpoint energy, $R_A=1.2A^{1/3}\,\text{fm}$, and $\tilde f_E$ a kinematic factor, see Ref.~\cite{companion} for details. In chiral EFT, all contributions in Eq.~\eqref{deltaNS0} are separately well defined, apart from the GT terms in which the scale- and scheme-dependent low-energy constants (LECs), labeled by $\tilde g_{Vp}^{N\!N}$, $\tilde g_{Vn}^{N\!N}$ in conventions in which they are expected to be of size ${\mathcal O}(1)$,
compensate an opposite dependence in $M^\text{mag}_{\text{GT}, N}$. The finite parts of the LECs  $\tilde g_{Vp}^{N\!N}$, $\tilde g_{Vn}^{N\!N}$ further encode effects of hard photons. 

The EFT matrix elements in Eq.~\eqref{deltaNS0} also capture a substantial part of the corrections that are traditionally modeled in the phase-space factor. In particular, $\overline{\delta^E_\text{NS}}$ reproduces the ${\mathcal O}(\alpha Z R)$ terms of the finite-size and shape corrections, while the ab initio $M^E_{\text{F},N}$ and $M^{+\,\text{3b}}_{\text{F},N}$ account for the different extent of the weak and charge distributions and for the nuclear-size dependence of the Fermi function, corrections that in traditional approaches were modeled, see Ref.~\cite{companion} for more details.

We evaluated the required NMEs in the coupled cluster (CC) method~\cite{Coester:1958seh, Coester:1960anj,Cizek:1966zzp,Kummel:1978zz, Bartlett:2007zza,Hagen:2013nca}, which is an ab initio many-body technique able to cover both the even--even~\cite{Ring:1980,Hagen:2007ew, Roth:2011vt, Duguet:2014jja, Qiu:2017, Tsuchimochi:2018, Novario:2020kuf} and odd--odd isotopes~\cite{Bartlett:2007zza, Shavitt:2009, Jansen:2011gb, Jansen:2012ey, Hagen:2012fb, Binder:2012mk} across the entire range from $^{10}$C $\rightarrow$ $^{10}$B to $^{54}$Co $\rightarrow$ $^{54}$Fe relevant for the $V_{ud}$ determination. Details of the evaluation are described in Ref.~\cite{companion}, with key points being that both truncation uncertainties in the many-body method and Hamiltonian uncertainties from the nuclear interaction can be estimated. For the latter, we compare calculations
using the
\intem~\cite{Entem:2003ft,Hebeler:2010xb} and \intsat~\cite{Ekstrom:2015rta} interactions. In addition, we performed benchmarks for light nuclei using
Quantum Monte Carlo (QMC) calculations for $^{14}$O $\rightarrow$ $^{14}$N~\cite{Cirigliano:2024msg} and $^{10}$C $\rightarrow$ $^{10}$B~\cite{King:2025fph,Graham3N}, suggesting that in the lightest system, due to $\alpha$-like clustering in the $^{10}$C wave function, the CC results might be less reliable, motivating $V_{ud}$ fits excluding this transition below.

\begin{figure*}[t]
 \centering
\includegraphics[width=0.9\linewidth]{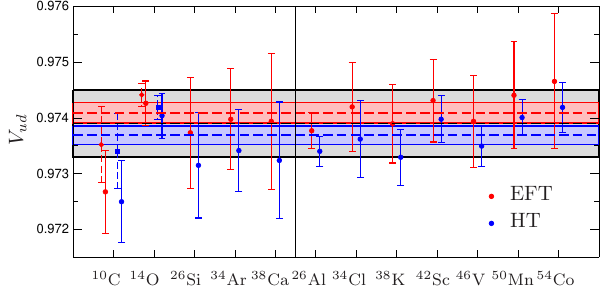}
\caption{$V_{ud}$ determined from each transition when setting the LECs to zero, comparing the EFT (red) and HT (blue) variants as described in the main text. The respective bands refer to the ``EFT (0)'' and ``HT (0)'' fits from Table~\ref{tab:fitVudLEC}, the gray band to the final value~\eqref{Vud_0+}, which includes an estimate of the uncertainty due to the LECs. The (solid) points indicated by diamonds are obtained using CC NMEs, the (dashed) points indicated by squares using QMC NMEs.}
\label{fig:Vud}
\end{figure*}

A special role is played by the three-body ${\mathcal O}(\alpha^2)$ NMEs $M^{+\, \text{3b}}_{\text{F}, N}$, which go beyond our CC campaign. Instead, we developed a formalism based on charge and EW densities that allows for an efficient estimate of such Fermi-type NMEs, and validated the results against the explicit ab initio calculations in the two-body sector as well as against QMC calculations for $^{10}$C and $^{14}$O in the three-body case. These NMEs prove important, despite the ${\mathcal O}(\alpha^2)$ and three-body suppression, because of enhancement by a large numerical factor and a coherent scaling with $Z^2$, the latter in contrast to the two-body analog $M^{+}_{\text{F}, N}$, in which case a cancellation between long- and short-distance contributions prevents any coherence. Further,  the density formalism used for the three-body NMEs allows us to verify that even higher-body NMEs are sufficiently suppressed.

Finally, the CC charge and EW densities enable further validation against experimental input. From the latter, we can derive weak transition radii $R_W$, which allow for complementary evaluations of the phase-space factor, thus providing a consistency check on the corrections between charge and EW distributions that enter in the standard formalism~\cite{Hardy:2020qwl,Hayen:2017pwg}. The charge densities imply predictions for charge radii, for which phenomenological input is available~\cite{Ohayon:2024dwt,Angeli:2013epw}. While neither interaction reproduces charge radii exactly, with \intsat~and \intem~leading to deviations $\simeq \pm 5\%$, respectively, the physical values are included in the range spanned by the two interactions, suggesting that the comparison should indeed provide a realistic estimate of the Hamiltonian uncertainties.

\begin{table*}[t]
\centering
    \renewcommand{\arraystretch}{1.3}
\begin{tabular}{c rrrr}
\toprule
Fit & \multicolumn{1}{c}{$V_{ud}$} & \multicolumn{1}{c}{$\tilde g_{Vp}^{{N\!N}}$} & \multicolumn{1}{c}{$\tilde g_{Vn}^{{N\!N}}$} & $\chi^2/\text{dof}$ \\\colrule
EFT (2) &    $0.97359(171)_{\text{stat}}
(12)_{g_V}(14)_{\mu} (105)_{\text{CC}}(2)_{\chi}$ & $2.2(4.8)_{\text{stat}}(0.3)_\mu (3.8)_{\text{CC}}(0.2)_{\chi}$ &
$-0.8(1.3)_{\text{stat}}(0.2)_{\mu}(0.5)_{\text{CC}}(0.3)_{\chi}$
& 0.2\\
EFT (1) &    $0.97390(21)_{\text{stat}}
(12)_{g_V}(3)_{\mu} (10)_{\text{CC}}(3)_{\chi}$ & -- &
$-1.0(1.0)_{\text{stat}}(0.1)_{\mu}(0.8)_{\text{CC}}(0.3)_{\chi}$
& 0.2\\
EFT (0) &    $0.97409(13)_{\text{stat}}
(12)_{g_V}(1)_{\mu} (3)_{\text{CC}}(4)_{\chi}$ & -- &
--
& 0.4\\\colrule
HT (2) &    $0.97267(83)_{\text{stat}}
(12)_{g_V}(40)_{\text{CC}}(12)_{\chi}$ & $3.7(2.4)_{\text{stat}} (1.7)_{\text{CC}}(0.3)_{\chi}$ &
$-0.7(1.1)_{\text{stat}}(0.3)_{\text{CC}}(0.3)_{\chi}$
& 0.8\\
HT (1) &    $0.97355(14)_{\text{stat}}
(12)_{g_V} (10)_{\text{CC}}(2)_{\chi}$ & -- &
$-1.2(9)_{\text{stat}}(8)_{\text{CC}}(2)_{\chi}$
& 0.9\\
HT (0) &    $0.97369(10)_{\text{stat}}
(12)_{g_V} (5)_{\text{CC}}(0)_{\chi}$ & -- &
--
& 1.2\\
\botrule
\end{tabular}
\caption{Fits to $V_{ud}$ and the LECs $\tilde g_{Vp}^{{N\!N}}$, $\tilde g_{Vn}^{{N\!N}}$ (in brackets the number of LECs included in the fit). The statistical uncertainties propagate the experimental errors and the uncertainties from the NMEs except for $M^\text{mag}_{\text{GT}, N}$ and $M^\text{CT}_{\text{GT}, N}$, $g_V$ indicates the universal single-nucleon uncertainties, $\mu$ the residual scale dependence in the EFT, CC the truncation uncertainty of the CC calculations of $M^\text{mag}_{\text{GT}, N}$ and $M^\text{CT}_{\text{GT}, N}$, and $\chi$ the Hamiltonian uncertainties.
}\label{tab:fitVudLEC}
\end{table*}

\emph{Global fit}---In general, the global fit to superallowed half-lives involves three free parameters, $V_{ud}$, $\tilde g_{Vp}^{N\!N}$, $\tilde g_{Vn}^{N\!N}$, since the two LECs are not predicted by chiral symmetry. Our fit is set up as follows:
\begin{enumerate}
 \item Scale uncertainties ($\mu$) and single-nucleon uncertainties ($g_V$) are treated as fully correlated.
 \item All NMEs except for $M^\text{mag}_{\text{GT}, N}$ and $M^\text{CT}_{\text{GT}, N}$ should not depend on the interaction, and we therefore assign uncertainties for each NMEs derived from the CC and Hamiltonian uncertainties as detailed in Ref.~\cite{companion}, which are added in quadrature to the experimental uncertainty, to define an overall ``statistical'' uncertainty component.
 \item For $M^\text{mag}_{\text{GT}, N}$ and $M^\text{CT}_{\text{GT}, N}$ we perform separate fits for each truncation and interaction, from which a final truncation (CC) and Hamiltonian ($\chi$) uncertainty is derived.
\end{enumerate}
Moreover, since not all ingredients of the master formula~\eqref{eq:master} are fully evaluated in the EFT yet, we explore different fit variants to assess the sensitivity to these non-EFT inputs. First, we consider fits that use Eq.~\eqref{deltaNS0} as is, thus using the maximal EFT input as currently possible, referred to as ``EFT'' in the following. Traditionally, some parts thereof have been included elsewhere, most notably the ${\mathcal O}(\alpha^2)$ and energy-dependent contributions as part fo the phase-space factor~\cite{Hardy:2020qwl}. Accordingly, we consider a ``HT'' fit variant, in which these terms are switched off in Eq.~\eqref{deltaNS0} and instead the input from Ref.~\cite{Hardy:2020qwl} is employed, see Ref.~\cite{companion} for details. In this way, the comparison of the two fits should provide an estimate of the impact of terms in the phase-space factor that are not evaluated fully consistently in the EFT yet.

As starting point, we consider the case in which the LECs are set to zero, so that a value of $V_{ud}$ follows directly from each transition, see Fig.~\ref{fig:Vud}. This figure illustrates that while there is in general excellent agreement among the different transitions, there is some tension for $^{10}$C when CC NMEs are employed, which reduces substantially in the case of QMC NMEs. Since for $^{14}$O the agreement between CC and QMC is indeed much better, this further supports the conclusion that the CC NMEs for $^{10}$C are less reliable. Once LECs are included, it is not possible to combine CC and QMC NMEs in a consistent manner, given that different nuclear interactions were employed, so that in view of the minor impact of $^{10}$C on the global fit, we concentrate here on variants in which this decay is excluded; see Ref.~\cite{companion} for fits that keep $^{10}$C, at the expense of a deteriorated $\chi^2$.

\begin{figure*}[t]
 \centering
\includegraphics[width=0.45\linewidth]{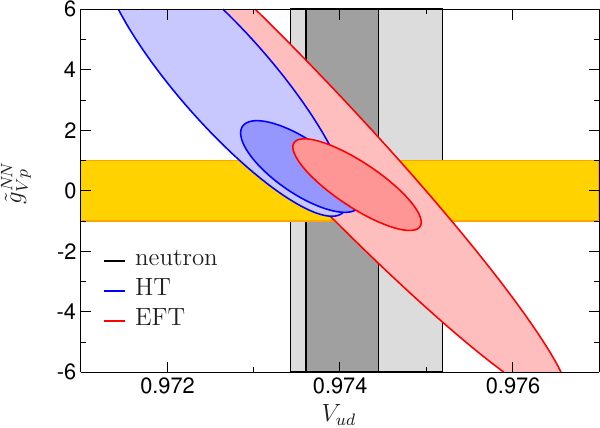}\quad
\includegraphics[width=0.45\linewidth]{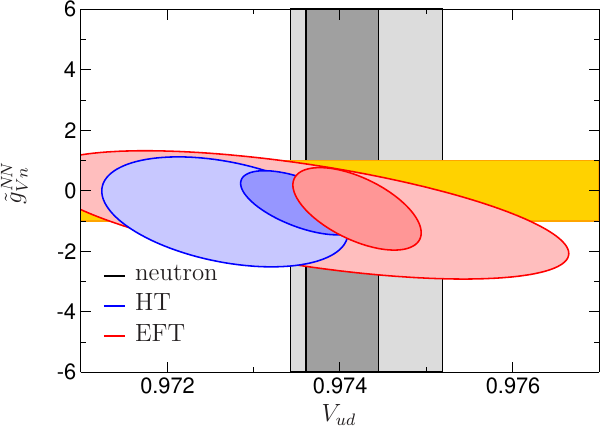}
\caption{Contours ($68\%$ confidence level) for $\tilde g_{Vp}^{N\!N}$ and $\tilde g_{Vn}^{N\!N}$ vs.\ $V_{ud}$ for the two-LEC fits, for the EFT (red) and HT (blue) versions. The light-shaded ellipses  refer to the fits (2) as given in Table~\ref{tab:fitVudLEC}, the darker-shaded ellipses to fits in which a Gaussian prior with width $1$ is imposed on the LECs. The orange band indicates the range $[-1,1]$, the gray bands $V_{ud}$ from neutron decay (dark gray as in Eq.~\eqref{Vud_neutron}, light gray including scale factors). }
\label{fig:Vudgvpgvn}
\end{figure*}

The results of the fits are shown in Table~\ref{tab:fitVudLEC}. In addition to the fits with LECs both set to zero or both included, we also consider a variant in which $\tilde g_{Vp}^{{N\!N}} = \frac{1+\kappa_p}{\kappa_n} \tilde g_{Vn}^{{N\!N}}$ is assumed, as motivated by the RG of the LECs, and only $\tilde g_{Vn}^{{N\!N}}$ retained as a fit parameter. From these fits, we observe:
\begin{enumerate}
 \item The central values for $V_{ud}$ in  the 0 LEC and 1 LEC fits come out close to Eq.~\eqref{Vud_HT}, and the difference between the EFT and HT fits is $\lesssim 4\times 10^{-4}$.
 \item In the 2 LEC fits the central value of $V_{ud}$ decreases substantially, accompanied by a sizable statistical error and central values
 $\tilde g_{Vp}^{{N\!N}} \gtrsim 1$. $V_{ud}$ and  $\tilde g_{Vp}^{{N\!N}}$ are strongly correlated, with a correlation coefficient $\simeq -0.97$.
\end{enumerate}
Accordingly, it is not possible to determine both LECs from a global fit, since $\tilde g_{Vp}^{{N\!N}}$ is too degenerate with $V_{ud}$, as illustrated in Fig.~\ref{fig:Vudgvpgvn}. The figure also shows the fit results when imposing a Gaussian prior with width $1$. In either case, the EFT variant of the fit leads to central values relatively close to the fits with zero or one LEC, while the HT variant implies a pull towards smaller values of $V_{ud}$.
The comparison to $V_{ud}$ from neutron decay further supports the indications for ${\mathcal O}(1)$ LECs from the more constrained fit variants, but scanning over the LECs within the range $\tilde g_{Vp}^{{N\!N}}, \tilde g_{Vn}^{{N\!N}}\in[-1,1]$ still suggests an uncertainty $\delta V_{ud}\simeq 6\times 10^{-4}$ that would need to be added to the zero-LEC results. In this way, we conclude that, accounting for the nuclear uncertainties following the EFT paradigm as far as possible, one obtains
\beq
\label{Vud_0+}
V_{ud}^{0+}=0.9739(6)\,,
\eeq
where the central value is defined by the average of EFT and HT zero-LEC fits.
Such a value for $V_{ud}$ from superallowed $\beta$ decays, combined with Eq.~\eqref{Vud_neutron} and the kaon constraints for $V_{us}$ as given in Ref.~\cite{Cirigliano:2022yyo}, reduces the unitarity deficit to $1.8\sigma$,  not because of a sizable shift in central value, but because of a more realistic estimate of the nuclear uncertainties.

\emph{Future prospects}---Comparing our current EFT evaluation~\eqref{Vud_0+} to the experimental plus single-nucleon uncertainty $\lesssim 2\times 10^{-4}$, there is a clear opportunity to improve the precision in $V_{ud}$ by a factor of three if the current limitations can be resolved. Most importantly, one needs independent input on $\tilde g_{Vp}^{{N\!N}}$, which in the traditional approach is implicit in the nuclear model, but, crucially, with uncertainties that are difficult to assess. In principle, information on the two-nucleon LECs is included in the dispersive approach of Refs.~\cite{Seng:2022cnq,Gennari:2024sbn}, in which the corresponding short-distance contributions are not parameterized explicitly, but are fixed implicitly by the modeling of the nuclear response at intermediate energies. Our analysis shows that $V_{ud}$ is highly sensitive to this information, so that a quantitative estimate of the associated uncertainty is essential to demonstrate control over the critical two-nucleon short-distance physics.
Matching the dispersive evaluation onto the EFT, by determining the effective $\tilde g_{Vp}^{{N\!N}}$, $\tilde g_{Vn}^{{N\!N}}$ it implies, would provide a direct cross-check of both approaches.

In this regard, the EFT offers the opportunity to again exploit the scale separation, and evaluate the LECs in the simpler two-nucleon system. Such an analysis could be performed in lattice QCD or using the Cottingham approach, as successfully demonstrated in the case of contact-term contributions to neutrinoless double $\beta$ decay~\cite{Cirigliano:2020dmx,Cirigliano:2021qko}.
Once the leading-order LECs are better determined, also uncertainties from higher orders in the chiral expansion need to be considered~\cite{Graham3N}.
Work along all these lines is in progress.

Besides the LECs, the isospin-breaking corrections represented by $\bar \delta_C$ need to be evaluated in an ab initio approach, and some early work in this direction has been performed in Refs.~\cite{Seng:2023cvt,Farren:2024spl,Piarulli:2026abq,Seng:2026hwk}. In particular, an ab initio evaluation of $\bar \delta_C$ requires one to keep track of isospin-breaking effects included in the nuclear interaction vs.\ those booked in the NMEs, such as the ${\mathcal O}(\alpha^2)$ contributions in Eq.~\eqref{deltaNS0} and the phase-space factor. This aspect emphasizes again the importance of using consistent interactions in different parts of the calculation, as is also crucial when nuclear-structure results from different methods, which might be better suited to certain isotopes, are to be combined. As illustrated here in the case of $^{10}$C, such combinations are only possible if consistent interactions have been employed, to the extent that one could only make use of the fact that Fig.~\ref{fig:Vud} clearly supports use of the QMC NMEs for $^{10}$C at the expense of an uncontrolled regulator dependence. 

Finally, the evaluation of the phase-space factor should be better controlled within the EFT, which requires a more detailed consideration of the current non-EFT inputs, see Refs.~\cite{companion,He:2026zie}. To this end, precision determinations of charge and EW densities would be valuable, which should at the same time help improve $\bar\delta_C$. While precision ab initio calculations are challenging given Hamiltonian uncertainties, such bulk properties have been found to be strongly correlated ~\cite{Hagen:2015yea,Payne:2019wvy,Heinz:2024cwg,Noel:2026byc}. This allows for precise evaluations as long as charge radii are available as calibration, which is the case for the isotopes relevant for superallowed $\beta$ decays~\cite{Ohayon:2024dwt}, and ab initio calculations for a sufficiently large sample of representative Hamiltonians are affordable. Needless to say, any additional experimental input to constrain the phase-space factors would also be highly welcome~\cite{Seng:2022inj,Gorchtein:2025wli}.

\emph{Conclusions}---In this Letter we reported substantial progress towards a complete EFT determination of $V_{ud}$ from superallowed $\beta$ decays. As key results, we calculated the NMEs required for $\bar\delta_\text{NS}$ for all relevant isotopes from  $^{10}$C $\rightarrow$ $^{10}$B to $^{54}$Co $\rightarrow$ $^{54}$Fe using the CC method, including detailed uncertainty estimates for truncations and chiral interactions. The NMEs were further validated against QMC calculations in light nuclei and estimates using nuclear densities, which were subsequently used as input for three-body matrix elements. The global fit to superallowed half-lives reveals that one LEC becomes almost degenerate with $V_{ud}$, requiring independent input and currently limiting the precision to $\delta V_{ud}\simeq 6\times 10^{-4}$, which, for the first time, is supported by an EFT treatment of $\bar\delta_\text{NS}$. We consider this work as the first step towards a full EFT determination of $V_{ud}$ from superallowed $\beta$ decays in the future, and formulated a roadmap of the next steps to ultimately realize a threefold reduction in the current $V_{ud}$ uncertainty.

\begin{acknowledgments}
\emph{Acknowledgments}---We thank Heiko Hergert, 
Garrett King, and Saori Pastore for helpful discussions. 
 Financial support by the  SNSF (Project No.\ TMCG-2\_213690) 
and the U.S.\ DOE (Grant No.\ DE-FG02-00ER41132) is gratefully acknowledged.
We acknowledge support from the DOE Topical Collaboration ``Nuclear Theory
for New Physics,'' Award No.\ DE-SC0023663. 
 Computer time was provided by the Innovative and Novel Computational Impact on Theory and Experiment (INCITE) program. This research used resources of the Oak Ridge Leadership Computing Facility located at ORNL, which is supported by the Office of Science of the Department of Energy under Contract No.~DE-AC05-00OR22725.  J.~d.~V.\ is funded by the Dutch Research Council (NWO) in the form of a VIDI grant and by the European Union (ERC, CRUNS, 101230525). Views and opinions expressed are, however,  those of the author(s) only and do not necessarily reflect those of the European Union or the European Research Council. Neither the European Union nor the granting authority can be held responsible for them.
 The work of S.G.\ and E.M.\ is supported by the U.S.\ Department of Energy through
the Los Alamos National Laboratory. Los Alamos National Laboratory is
operated by Triad National Security, LLC, for the National Nuclear
Security Administration of U.S.\ Department of Energy (Contract No.\
89233218CNA000001), by the Office of Advanced Scientific Computing
Research, Scientific Discovery through Advanced Computing (SciDAC)
NUCLEI program, and by the LANL LDRD program, under projects
20260246ER and 20270410ER.
This research used resources provided by the Los Alamos National
Laboratory Institutional Computing Program, which is supported by the
U.S.\ Department of Energy National Nuclear Security Administration under
Contract No.\ 89233218CNA000001. The work of G.C.\ was supported by the U.S.\ Department of Energy (No.\ DE-SC0021027). QMC calculations were performed by the National Energy Research Scientific Computing Center under Contract No.\ DE-AC02-05CH11231
using NERSC award NP-ERCAP0036336 and NP-ERCAP0032885.
\end{acknowledgments}

\bibliography{delta}

\end{document}